\documentstyle[twoside,psfig]{article}

\catcode`\@=11
\long\def\@makefntext#1{
\protect\noindent \hbox to 3.2pt {\hskip-.9pt  
$^{{\eightrm\@thefnmark}}$\hfil}#1\hfill}		%CAN BE USED 

\def\@makefnmark{\hbox to 0pt{$^{\@thefnmark}$\hss}}	%ORIGINAL 
	
\def\ps@myheadings{\let\@mkboth\@gobbletwo
\def\@oddhead{\hbox{}
\rightmark\hfil\eightrm\thepage}   
\def\@oddfoot{}\def\@evenhead{\eightrm\thepage\hfil
\leftmark\hbox{}}\def\@evenfoot{}
\def\sectionmark##1{}\def\subsectionmark##1{}}

\oddsidemargin=\evensidemargin
\newcounter{sectionc}\newcounter{subsectionc}\newcounter{subsubsectionc}
\renewcommand{\section}[1] {\vspace{12pt}\addtocounter{sectionc}{1} 
\setcounter{subsectionc}{0}\setcounter{subsubsectionc}{0}\noindent 
	{\tenbf\thesectionc. #1}\par\vspace{5pt}}
\renewcommand{\subsection}[1] {\vspace{12pt}\addtocounter{subsectionc}{1} 
	\setcounter{subsubsectionc}{0}\noindent 
	{\bf\thesectionc.\thesubsectionc. {\kern1pt \bfit #1}}\par\vspace{5pt}}
\renewcommand{\subsubsection}[1] {\vspace{12pt}\addtocounter{subsubsectionc}{1}
	\noindent{\tenrm\thesectionc.\thesubsectionc.\thesubsubsectionc.
	{\kern1pt \tenit #1}}\par\vspace{5pt}}
\newcommand{\nonumsection}[1] {\vspace{12pt}\noindent{\tenbf #1}
	\par\vspace{5pt}}

\newcounter{appendixc}
\newcounter{subappendixc}[appendixc]
\newcounter{subsubappendixc}[subappendixc]
\renewcommand{\thesubappendixc}{\Alph{appendixc}.\arabic{subappendixc}}
\renewcommand{\thesubsubappendixc}
	{\Alph{appendixc}.\arabic{subappendixc}.\arabic{subsubappendixc}}

\renewcommand{\appendix}[1] {\vspace{12pt}
        \refstepcounter{appendixc}
        \setcounter{figure}{0}
        \setcounter{table}{0}
        \setcounter{lemma}{0}
        \setcounter{theorem}{0}
        \setcounter{corollary}{0}
        \setcounter{definition}{0}
        \setcounter{equation}{0}
        \renewcommand{\thefigure}{\Alph{appendixc}.\arabic{figure}}
        \renewcommand{\thetable}{\Alph{appendixc}.\arabic{table}}
        \renewcommand{\theappendixc}{\Alph{appendixc}}
        \renewcommand{\thelemma}{\Alph{appendixc}.\arabic{lemma}}
        \renewcommand{\thetheorem}{\Alph{appendixc}.\arabic{theorem}}
        \renewcommand{\thedefinition}{\Alph{appendixc}.\arabic{definition}}
        \renewcommand{\thecorollary}{\Alph{appendixc}.\arabic{corollary}}
        \renewcommand{\theequation}{\Alph{appendixc}.\arabic{equation}}
        \noindent{\tenbf Appendix \theappendixc #1}\par\vspace{5pt}}
\newcommand{\subappendix}[1] {\vspace{12pt}
        \refstepcounter{subappendixc}
        \noindent{\bf Appendix \thesubappendixc. {\kern1pt \bfit #1}}
	\par\vspace{5pt}}
\newcommand{\subsubappendix}[1] {\vspace{12pt}
        \refstepcounter{subsubappendixc}
        \noindent{\rm Appendix \thesubsubappendixc. {\kern1pt \tenit #1}}
	\par\vspace{5pt}}

\newcommand{\textlineskip}{\baselineskip=13pt}
\newcommand{\smalllineskip}{\baselineskip=10pt}

\def\eightcirc{
\begin{picture}(0,0)
\put(4.4,1.8){\circle{6.5}}
\end{picture}}
\def\eightcopyright{\eightcirc\kern2.7pt\hbox{\eightrm c}} 

\newcommand{\copyrightheading}[1]
	{\vspace*{-2.5cm}\smalllineskip{\flushleft
	{\footnotesize Modern Physics Letters A, #1}\\
	{\footnotesize $\eightcopyright$\, World Scientific Publishing
	 Company}\\
	 }}

\newcommand{\bibit}{\nineit}

\renewenvironment{thebibliography}[1]
	{\frenchspacing
	 \ninerm\baselineskip=11pt
	 \begin{list}{\arabic{enumi}.}
        {\usecounter{enumi}\setlength{\parsep}{0pt}     
	 \setlength{\leftmargin 12.7pt}{\rightmargin 0pt} %FOR 1--9 ITEMS
         \setlength{\itemsep}{0pt} \settowidth
	{\labelwidth}{#1.}\sloppy}}{\end{list}}

\newcounter{itemlistc}
\newcounter{romanlistc}
\newcounter{alphlistc}
\newcounter{arabiclistc}

\newcommand{\fcaption}[1]{
        \refstepcounter{figure}
        \setbox\@tempboxa = \hbox{\footnotesize Fig.~\thefigure. #1}
        \ifdim \wd\@tempboxa > 5in
           {\begin{center}
        \parbox{5in}{\footnotesize\smalllineskip Fig.~\thefigure. #1}
            \end{center}}
        \else
             {\begin{center}
             {\footnotesize Fig.~\thefigure. #1}
              \end{center}}
        \fi}

\newcommand{\tcaption}[1]{
        \refstepcounter{table}
        \setbox\@tempboxa = \hbox{\footnotesize Table~\thetable. #1}
        \ifdim \wd\@tempboxa > 5in
           {\begin{center}
        \parbox{5in}{\footnotesize\smalllineskip Table~\thetable. #1}
            \end{center}}
        \else
             {\begin{center}
             {\footnotesize Table~\thetable. #1}
              \end{center}}
        \fi}

\def\@citex[#1]#2{\if@filesw\immediate\write\@auxout
	{\string\citation{#2}}\fi
\def\@citea{}\@cite{\@for\@citeb:=#2\do
	{\@citea\def\@citea{,}\@ifundefined
	{b@\@citeb}{{\bf ?}\@warning
	{Citation `\@citeb' on page \thepage \space undefined}}
	{\csname b@\@citeb\endcsname}}}{#1}}

\newif\if@cghi
\def\cite{\@cghitrue\@ifnextchar [{\@tempswatrue
	\@citex}{\@tempswafalse\@citex[]}}
\def\citelow{\@cghifalse\@ifnextchar [{\@tempswatrue
	\@citex}{\@tempswafalse\@citex[]}}
\def\@cite#1#2{{$\null^{#1}$\if@tempswa\typeout
	{IJCGA warning: optional citation argument 
	ignored: `#2'} \fi}}

\def\pmb#1{\setbox0=\hbox{#1}
	\kern-.025em\copy0\kern-\wd0
	\kern.05em\copy0\kern-\wd0
	\kern-.025em\raise.0433em\box0}

\def\fnt#1#2{\footnotetext{\kern-.3em
	{$^{\mbox{\scriptsize #1}}$}{#2}}}

\def\fpage#1{\begingroup
\voffset=.3in
\thispagestyle{empty}\begin{table}[b]\centerline{\footnotesize #1}
	\end{table}\endgroup}

\font\tenrm=cmr10
\font\tenit=cmti10 
\font\tenbf=cmbx10
\font\bfit=cmbxti10 at 10pt
\font\ninerm=cmr9
\font\nineit=cmti9

\font\eightrm=cmr8

\def\qed{\hbox{${\vcenter{\vbox{			%HOLLOW SQUARE
   \hrule height 0.4pt\hbox{\vrule width 0.4pt height 6pt
   \kern5pt\vrule width 0.4pt}\hrule height 0.4pt}}}$}}

\begin{document}
\setlength{\textheight}{7.7truein}  %for 2nd page onwards

%\runninghead{Instructions for Typesetting Camera-Ready
%Manuscripts $\ldots$}{Instructions for Typesetting Camera-Ready
%Manuscripts $\ldots$}

\normalsize\textlineskip
\thispagestyle{empty}
\setcounter{page}{1}

\copyrightheading{}			%{Vol. 0, No.0 (1992) 000--000}

\vspace*{0.88truein}

\fpage{1}
\begin{flushright}
{\bf{\normalsize{IC/00/45}}}
\end{flushright}
\vspace*{30pt}
\centerline{\bf Q--ball formation in the MSSM with explicit CP violation}
\vspace*{0.37truein}
\centerline{\footnotesize M. Boz}
\baselineskip=12pt
\centerline{\footnotesize\it Physics  Department, Hacettepe University}
\baselineskip=10pt
\centerline{\footnotesize\it Ankara, 06533, Turkey }
\vspace*{10pt}

\centerline{\footnotesize D. A. Demir}
\baselineskip=12pt
\centerline{\footnotesize\it The Abdus Salam International Center For
Theoretical Physics}
\baselineskip=10pt
\centerline{\footnotesize\it Trieste, I-34100, Italy }
\vspace*{10pt}

\centerline{\footnotesize N. K. Pak}
\baselineskip=12pt
\centerline{\footnotesize\it Physics Department, Middle East
Technical University}
\baselineskip=10pt
\centerline{\footnotesize\it Ankara, 06531,  Turkey }
\vspace*{0.225truein}
%%%%%%%%%%%%%%%%%%%%%%%%%%%%%%%%%%%%%%%%%%%%%%%%%%%%%%%%%%%%%%%%%%%%%%%%%
%%%%%%%%%%%%%%%%%%%%%%%%%%%%%%%%%%%%%%%%%%%%%%%%%%%%%%%%%%%%%%%%%%%%%%%%%
\begin{abstract}
\noindent
Q--balls generically exist in the supersymmetric extensions of the standard
model. Taking into account the additional sources of CP violation,
which are naturally accomodated by the supersymmetric models, it is  shown that the 
Q--ball matter depends additively on individual CP phases, whereas mass per unit charge in the
Q--ball depends only on the relative phases. There are regions of the parameter space 
where there is no stable Q--ball solution in the CP--conserving limit whereas finite 
CP phases induce a stable Q ball.
\end{abstract}
\section{Introduction}
Non-topological solitons, in particular Q--balls, are extended objects with finite mass and
spatial extension, and arise in scalar field theories when there is an exact
continious symmetry and some kind of attractive interaction, as already 
classified by Coleman \cite{coleman}.

As first pointed out by Kusenko \cite{kusenko}, Q-balls naturally exist in supersymmetric
theories thanks to global baryon (B) and lepton (L) number symmetries. Besides, theories
with an extended scalar sector can support non--baryonic Q--balls \cite{durmus}. Q--balls
have found applications in modelling several physical processes ranging from leptoquarks
to dark matter \cite{apply1,apply2}. 

It is well known that the MSSM has unremovable physical phases which can be identified 
with the phases of $\mu$ parameter and $A$ terms \cite{phases}. According to the vacuum
stability arguments these phases relax to CP conserving points \cite{dimo}. However, the
same arguments are not sufficient to relax the phases in the minimal extensions \cite{benim}.
Hence, it is plausable to take these phases finite and look for their effects in low energy 
processes. 

As summarized above, the MSSM predicts the existence of both non--topological
solitons and finite CP violation. However, so far the investigations on the Q--ball formation 
in the MSSM have not dealt with the effects of the CP violation. In this short note we investigate 
the effects of explicit CP violation in the MSSM on the Q--ball formation. In the next 
section we discuss this issue in detail using the MSSM scalar potential. We particularly analyze
the effects of the phases in the trilinear couplings and the $\mu$ parameter.  In the last section we 
summarize the main findings.   

\section{MSSM with explicit CP violation and Q-balls}
The MSSM scalar sector contains two Higgs doublets (with opposite hypercharge) 
and scalar partners
of quarks and leptons. The supersymmetry and gauge symmetry are broken by the soft supersymmetry
breaking terms which introduce a number of mass parameters to the potential. Both the Higgsino
Dirac mass term $\mu$ and the trilinear couplings in soft terms are complex, and they lead to 
CP violation beyond the CKM matrix already present  in the SM. Denoting the neutral components of the
Higgs doublets as $\phi_1$ and $\phi_2$, the MSSM scalar potential for one generation of sfermions
reads as: 
\begin{eqnarray}
V_{MSSM}&=&\big[(h_{u}A_{u}\phi_{2}-h_{u}\mu^{*}\phi_{1}^{*}) \tilde{u}_{L}\tilde{u}_{R}^{*}
-(h_{d}A_{d} \phi_{1}-h_{d}\mu^{*}\phi_{2}^{*})\tilde{d}_{L}\tilde{d}_{R}^{*}\nonumber\\
&&-(h_{e}A_{e} \phi_{1}-h_{e}\mu^{*}\phi_{2}^{*})\tilde{e}_{L}\tilde{e}_{R}^{*}+h.c.\big]\nonumber\\
&&+m_{\tilde{Q}}^{2}|\tilde{Q}|^{2}
+m_{\tilde{u}}^{2}|\tilde{u}_{R}|^{2}
+m_{\tilde{d}_{R}}^{2}|\tilde{d}_{R}|^{2}
+m_{\tilde{L}}^{2}|\tilde{L}|^{2}
+m_{\tilde{e}_{R}}^{2}|\tilde{e}_{R}|^{2}\nonumber\\
&&+m_{1}^{2}|\phi_{1}|^{2}
+m_{2}^{2}|\phi_{2}|^{2}
+|\mu|^{2}|\phi_{2}|^{2}+|\mu|^{2}|\phi_{1}|^{2}
+h_{e}^{2}|\tilde{L}^{2}||\tilde{e}_{R}|^{2}\nonumber\\
&&+h_{d}^{2}|\tilde{Q}^{2}||\tilde{d}_{R}|^{2}
+h_{e}^{2}|\tilde{L}^{2}||\tilde{e}_{R}|^{2}+h_{u}^{2}|\tilde{Q}^{2}
||\tilde{u}_{R}|^{2}+h_{e}^{2}|\phi_{1}|^{2}|\tilde{e}_{R}|^{2}\nonumber\\
&&+h_{d}^{2}|\phi_{1}|^{2}|\tilde{d}_{R}|^{2}+h_{e}^{2}|\phi_{1} \tilde{e}_{L}|^{2}+
h_{u}^{2}|\phi_{2}\tilde{u}_{L}|^{2}+h_{d}^{2}|\phi_{1}\tilde{d}_{L}|^{2}
\end{eqnarray}
plus the $D$ term contributions which will not be shown explicitely. This scalar potential
has two global symmetries: $U(1)_B$ and $U(1)_L$ corresponding to baryon number and lepton
number symmetries, respectively. These are the exact symmetries of the theory and their
breaking (spontaenous and otherwise) lead to B-- and L-- violating
processes. As was emphasized
in Refs. 1 and 2 it is mainly the cubic couplings that generate B--ball or L--ball 
type solitonic solutions. It is appearent that slepton doublet $\tilde{L}$ and right--handed 
slepton $\tilde{e}_{R}$ contribute to L--balls whereas one needs the squark doublet  $\tilde{Q}$
and right--handed squarks $\tilde{u}_{R}$ and $\tilde{d}_{R}$ to form B--balls. In both cases
Higgs fields are necessary. In this form the scalar potential involves
several scalar fields,
and a true analysis of the Q--ball formation requires a minimization of the multi--field quantity
\cite{coleman}
\begin{eqnarray}
\label{massper}
m_{eff}^{2}(\phi_1,\cdots,\tilde{e}_R)\equiv 2 V_{MSSM}/\sum_{Ball} charge\times |field|^{2}
\end{eqnarray} 
that guarantees the stability of the Q--ball against decaying into its constitutents. In this
formula $charge$ and $field$ denote the baryon number (lepton number) and squark fields (slepton 
fields). Instead of dealing with coupled equations of motion for fields contributing to a 
particular Q--ball, practically one can describe the nature of the Q--ball by using a single scalar
degree of freedom \cite{kusenko}. This approximation is quite accurate especially for $D$-- and
$F$--flat potentials as the degrees of freedom orthogonal to  flat directions will be much more massive
\cite{apply1,apply2}. For this purpose it is convenient to introduce a scalar field $\varphi$
representing the
Q-ball matter, and decompose the component fields in terms of $\varphi$ using the dimensionless
parameters $\xi_i$, with $\xi<1$ and $\sum_{i}{\xi_{i}^{2}}=1$, as follows: 
\begin{eqnarray}  
\tilde{e}_{L,R}=\xi_{e_{L,R}}\varphi,\ \ \ \tilde{u}_{L,R}=\xi_{u_{L,R}}\varphi,\ \ \
\tilde{d}_{L,R}=\xi_{d_{L,R}}\varphi,\ \ \ \ \phi_{1,2}=\xi_{1,2}\varphi~.
\end{eqnarray}
Using this decomposition, the scalar potential $V_{MSSM}$ takes the form  
\begin{eqnarray}
\label{LBSP}
V_{\varphi}=M_{\varphi}^{2}|\varphi|^{2}
+M_{c} Re \varphi|\varphi|^{2}
-M_{s} Im \varphi|\varphi|^{2}+\lambda|\varphi|^{4}
\end{eqnarray}
for both $L$-- and $B$--balls. Here $M_{\varphi}^{2}$ is the linear combination of the scalar 
quadratic mass parameters in $V_{MSSM}$. The quartic coupling $\lambda$ is a linear combination
of Yukawa couplings $h_{u,d,e}$ and gauge couplings $g_{3,2,1}$ following, respectively, from the
$F$--term and $D$--term contributions. 

As the general analyses of Q--ball formation Refs. 1, 2, 3 show explicitly, the 
crucial parameters in Eq. (\ref{LBSP}) are the trilinear mass parameters $M_c$ and $M_s$. For $L$--balls
one has 
%\cite{coleman,kusenko,durmus}
\begin{eqnarray}
\label{Ml}
M_{c}&=&2\left[-\cos(\phi_{A_{e}})\tilde{A}_{e}+\cos(\phi_{\mu})\tilde{\mu}\right]\\
M_{s}&=&2\left[-\sin(\phi_{A_{e}})\tilde{A}_{e}+\sin(\phi_{\mu})\tilde{\mu}\right]
\end{eqnarray} 
where
\begin{eqnarray}
\label{ae}
\tilde{A_{e}}&=&h_{e}|A_{e}|\xi_{1}\xi_{e_{L}}\xi_{e_R}\nonumber\\
\tilde{\mu}&=&h_{e}|\mu|\xi_{2}\xi_{e_L}\xi_{e_R}
\end{eqnarray}
where $\phi_{A_{e}}$ and $\phi_{\mu}$ are the phases of $A_e$ and $\mu$ parameter, respectively.

For $B$--balls, however, $M_c$ and $M_s$ have the following expressions
\begin{eqnarray}
\label{MB}
M_{c}&=&2\left[\cos(\phi_{A_{u}})\tilde{A_{u}}
-\cos(\phi_{A_{d}})\tilde{A_{d}}+\cos(\phi_{\mu})\tilde{\mu}\right]\\
M_{s}&=&2\left[\sin(\phi_{A_{u}})\tilde{A_{u}}-\sin(\phi_{A_{d}})\tilde{A_{d}}+
\sin(\phi_{\mu})\tilde{\mu}\right]
\end{eqnarray}
where 
\begin{eqnarray}
\label{ae2}
\tilde{A_{u}}&=&h_{u}|A_{u}|\xi_{2}\xi_{u_{L}}\xi_{u_{R}}\nonumber\\ 
\tilde{A_{d}}&=&h_{d}|A_{d}|\xi_{1}\xi_{d_{L}}\xi_{d_{R}}\nonumber\\
\tilde{\mu}&=&|\mu|(h_d \xi_{2}\xi_{d_{L}}\xi_{d_{R}}-h_u \xi_{1}\xi_{u_{L}}\xi_{u_{R}})~.
\end{eqnarray}

The form of the scalar potential Eq. (\ref{LBSP}) is such that $U(1)_B$ or $U(1)_L$ symmetries are not
manifest at all. In particular, $Im \varphi$ and $Re \varphi$ refer to Higgs fields $\phi_{1,2}$ which 
do not contribute to the charge of the Q--ball. For instance, in thin--wall 
approximation and in the notation of Ref. 2, the total
charge and the effective mass of the $B$--ball are respectively given by 
\begin{eqnarray}
\label{charge}
B&=& 2a\omega |\varphi|^{2} V\nonumber\\
m_{eff}^{2}&=&\frac{1}{a}\left[ M_{\varphi}^{2}
+M_{c} Re \varphi
-M_{s} Im \varphi+\lambda\{( Re \varphi)^{2}
+( Im \varphi)^{2}\}\right] \nonumber\\
\rm{where}\,\,\,\, 
a&=&\frac{1}{3}\left[\xi_{u_{L}}^{2}+\xi_{u_{R}}^{2}+\xi_{d_{L}}^{2}+\xi_{d_{R}}^{2}\right] 
\end{eqnarray}
 Thus, the total charge vanishes if squarks are absent; that is, 
the Higgs fields do not play a role in charge accumulation. However, there is no stable Q--ball  if 
the Higgs fields are absent, as can be seen from vanishing of the trilinear couplings $M_c$ and
$M_s$. In this sense, $Re$ and $Im$ parts of $\varphi$ in Eq. (\ref{LBSP}) refer to the time--independent 
phase of $\varphi$ generated by the non--trivial phases in $\mu$ and $A_{u,d,e}$ parameters. 

As is seen from Eq. (\ref{LBSP}), the main effect of complex $\mu$ and $A_{u,d,e}$ is to introduce
$M_s\neq 0$ which is proportional to $Im \varphi$. There are three distinct limits in which the
Q--ball matter gains different CP characteristics depending on the values of $M_c$
and $M_s$.  

So far discussions of Q--balls have been based on purely real $\mu$ and $A_{u,d,e}$ in which 
case $M_s\equiv 0$. Then only $Re \varphi$ has a trilinear coupling and stable Q--matter is thus composed 
of $Re \varphi$ which is purely CP even. 

In the opposite limit, that is, for purely imaginary $\mu$ and $A_{u,d,e}$ one has $M_c\equiv 0$. Then only
$Im \varphi$ has a trilinear coupling, and thus, the resulting Q--ball is
composed of $Im \varphi$, and Q--matter is purely CP odd. 

In the general case, where $\mu$ and $A_{u,d,e}$ are complex parameters with nonvanishing real and
imaginary parts, both $Im \varphi$ and $Re \varphi$ contribute to the Q--matter. Namely, $m_{eff}^{2}$ is
minimized for $Re \varphi = -M_c/2\lambda$ and $Im \varphi = M_s/2\lambda$, so that mass per unit charge 
for $B$--ball reads as 
\begin{eqnarray} 
\label{mef2}
&&m_{eff}^{2}(B-ball)=\frac{1}{a}\left[M_{\varphi}^{2}-
\frac{1}{4 \lambda}( M_{c}^{2} +M_{s}^{2})\right]~.
\end{eqnarray}
According to Coleman's theorem \cite{coleman}, if $0< m_{eff}^{2} < M_{\varphi}^{2}/a$ then the resulting B-ball is stable. 
One also notices that  $m_{eff}^{2}$ depends explicitly on the relative phase
between any pair of $\tilde{A}_{d}, \tilde{A}_{u}$ and $\tilde{\mu}$, using
Eq.(\ref{MB}):
\begin{eqnarray} 
&&m_{eff}^{2}(B-ball)=\frac{1}{a}\Bigg[M_{\varphi}^{2}-\frac{1}{\lambda}\left( \tilde{A_{u}}^{2}+
\tilde{A_{d}}^{2}+
\tilde{\mu}^{2}-2 \tilde{A_{u}} \tilde{A_{d}} \cos (\phi_{A_u}-\phi_{A_d})\right.\nonumber\\
&&\left.+2\tilde{A_{u}} \tilde{\mu}
\cos (\phi_{A_u}-\phi_{\mu}) - 2 \tilde{A_{d}} \tilde{\mu}\cos
(\phi_{A_d}-\phi_{\mu})\right)\Bigg]
\end{eqnarray}
Depending on the values of these CP phases $m_{eff}^{2}$ takes on a range of values. Therefore, 
mass per unit charge in the $B$--ball varies with the CP violating phases.
To illustrate this, one can consider the
simple case of $\tilde{u}_L\tilde{u}_R$ $B$--ball, that is, $A_d=0$. In this case $m_{eff}^{2}$ varies
from its minimum value ($\phi_{A_u}-\phi_{\mu}=0$)
\begin{eqnarray} 
\left[m_{eff}^{2}(B-ball)\right]_{min}=\frac{1}{a}\Big[M_{\varphi}^{2}-\frac{1}
{\lambda}\left(
\tilde{A_{u}}+\tilde{\mu}\right)^{2}\Big]
\end{eqnarray}
to the maximal value ($\phi_{A_u}-\phi_{\mu}=\pi$) 
\begin{eqnarray}
\left[m_{eff}^{2}(B-ball)\right]_{max}=\frac{1}{a}\Big[M_{\varphi}^{2}-\frac{1}{\lambda}\left(\tilde{A_{u}}-
\tilde{\mu}\right)^{2}\Big]
\end{eqnarray}
with the mean ($\phi_{A_u}-\phi_{\mu}=\pi/2$) 
\begin{eqnarray}
\left[m_{eff}^{2}(B-ball)\right]_{mean}=\frac{1}{a}\Big[M_{\varphi}^{2}-\frac{1}{\lambda}\left(\tilde{A_{u}}^{2}+
\tilde{\mu}^{2}\right)\Big]
\end{eqnarray}
taking $\tilde{\mu}$ positive. One notices that none of the conditions above implies a specific value 
for the phases $\phi_{A_u}$ and $\phi_{\mu}$. Indeed the mass parameters $M_c$ and $M_s$ take on the following values 
\begin{eqnarray}
\label{minimax}
&&M_c=2\left(\tilde{A_{u}}+\tilde{\mu}\right)~ \cos \phi_{A_u}~,\ \ M_s=2\left(\tilde{A_{u}}+\tilde{\mu}\right)~ \sin \phi_{A_u}\nonumber\\
&&M_c=2\left(\tilde{A_{u}}-\tilde{\mu}\right)~ \cos \phi_{A_u}~,\ \ M_s=2\left(\tilde{A_{u}}-\tilde{\mu}\right)~ \sin \phi_{A_u}\\
&&M_c=2\left(\tilde{A_{u}}\cos \phi_{A_u}+\tilde{\mu}\sin \phi_{A_u}\right)~,\ \ M_s=2\left(\tilde{A_{u}}\sin \phi_{A_u}-\tilde{\mu}\cos
\phi_{A_u}\right)\nonumber
\end{eqnarray}
for $\phi_{A_u}-\phi_{\mu}=0$, $\pi$ and $\pi/2$, respectively. Hence, despite the phase independence of mass per unit charge (14)--(16), 
the value of the condensate, determined by $M_c$ and $M_s$,  is an explicit function of the CP phases.  This follows from the fact that
the Q--matter $\varphi$ depends on the individual phases additively whereas $m_{eff}^{2}$ depends only on the relative phases. This
particular pattern of phase structure can be important for Q--ball formation. In the purely CP--conserving limit, one would have 
$m_{eff}^{2}=\left[m_{eff}^{2}(B-ball)\right]_{min}$ together with $M_c=2\left(\tilde{A_{u}}+\tilde{\mu}\right)$ and $M_s=0$, 
correspondig to the first line of (\ref{minimax}) with $\phi_{A_u}=0$. However, as the Eqs. (14)--(16) and (\ref{minimax}) suggest
clearly the nonvanishing CP phases offer more alternatives.

To see the implications of such a phase dependence, one can consider the special case of $\tilde{A_{u}}=\tilde{\mu}$ and 
$\tilde{\mu}^{2}/\lambda=M_{\varphi}^{2}/4$. Then, for $\phi_{A_u}-\phi_{\mu}=0$, one has $m_{eff}^{2}=0$ with
nonvanishing $M_c$ and $M_s$. Therefore, for this parameter set one has a massless Q--ball, or equivalently, a Q--matter 
distribution over entire space. In this case, at least in the thin--wall approximation, there is no macroscopic structure
with finite size. On the other hand, for $\phi_{A_u}-\phi_{\mu}=\pi$,  $m_{eff}^{2}= M^{2}_{\varphi}/a$, and this corresponds to
the critical value of $m_{eff}^{2}$ below  which there would be a stable Q--ball solution. A careful look at $M_c$ and $M_s$ shows
that they vanish identically and Q--ball formation without the trilinear couplings is already out of question. Therefore this
possibility leaves no room for Q--ball formation. Finally, for $\phi_{A_u}-\phi_{\mu}=\pi/2$, however, one obtaines
$m_{eff}^{2}=M^{2}_{\varphi}/2 a$ leaving both $M_c$ and $M_s$ nonvanishing. This is a regular Q--ball solution, and it
corresponds to maximal CP--violation, for instance, in the Higgs sector \cite{phases}. Indeed, $M_c$ and $M_s$ can vary with
$\phi_{A_u}$ further and this does not affect the Q--ball structure obtained for $\left[m_{eff}^{2}(B-ball)\right]_{mean}$.
Therefore, depending on the amount of CP violation, there may be regions of the parameter space that support a stable Q--ball
solution though the strictly CP--conserving  MSSM does not. 
 
Another implication of these phases would be on the scattering of fermions from the Q--ball. Indeed, a typical cross section 
has the form $\sigma\sim 4 \pi R^{2}$ where $R^{2}\sim 1/m_{eff}^{2}$. Therefore, the supersymmetric CP phases influence the 
formation as well as interactions of the Q--balls with surrounding plasma. In fact, this expectation is confirmed by the 
recent analysis of the scattering of the dark matter particles from the nucleons \cite{dark}.

So far we have discussed Q--ball formation without imposing any $D$-- and/or $F$--flatness. However,  
the scalar potential of the low--energy supersymmetric theories has many flat directions. Such flat   
potentials are phenomenologically relevant as it is possible to produce large enough Q--balls that can
resist the evaporation on time scales of the order of the age of the universe \cite{apply1,apply2}. As
has been listed in Ref. [9], there are slepton as well as squark flat directions with corresponding
Q--balls. In fact, radiatively corrected flat directions in the MSSM induce
a potential of the form
\begin{eqnarray}
V_{\varphi}=M_{\varphi}^{2}|\varphi|^{2}+\lambda^{2} \frac{|\varphi|^{2(d-1)}}{M_{Pl}^{2(d-3)}}
+\left(\lambda A \frac{\varphi^{d}}{M_{Pl}^{(d-3)}} + h. c.\right)
\end{eqnarray}
where $d$ is the mass dimension of the nonrenormalizable operator in the potential, and $A$ is a typical
trilinear coupling in the soft supersymmetry breaking part. For the purpose of this work the essential
piece in this formula is $A$ dependent part. As discussed above, if $A$--terms are complex
parameters, in general, the $Re \varphi$ as well as $Im \varphi$ can develop nonvanishing vacuum
expectation values leading to CP violating Q--matter.

\section{Discussions}
In this work we have discussed Q-ball formation in the MSSM with explicit CP violation. It is seen that
the complex $\mu$ parameter and $A$--terms can affect the Q-ball formation process. In particular, it is
shown that the scalar vacuum expectation value in the Q--ball depends additively on the soft CP phases
whereas the mass per unit charge of the Q--ball depends only on the relative phases.
There are regions of the parameter space where finite CP phases induce a stable Q--ball where 
the strictly CP--conserving does not. 

Once the Q--ball is formed, as long as the Q--matter is CP violating one, one expects that the
scatterings of the light fermions from the Q--ball as well as its decay to fermions (neutralino LSP, for
example) can affect their phenomenology. Particularly interesting is the Q--matter contributing to dark
matter where the detection rates will change with the CP violating phases considerably \cite{dark}.

It is worthy of reemhasizing that:
\begin{enumerate}
\item The vacuum expectation values of the fields, that is, the Q--ball matter $\varphi$
depends on the individual relative CP phases additively; however,
\item The mass per unit charge in the Q-ball
depends only on the relative phases.
\end{enumerate}
Hence, there are regions of the parameter space where the Q--matter
is nonvanishing; however, the Q--ball solutions are not stable. Moreover, there are regions of the
parameter space where finite CP phases suport a stable Q--ball whereas the CP conserving MSSM does not.
These discussions equally apply to $L$--balls as well, and one may obtain new kinds of Q--ball solutions 
by including appropriate soft terms such as $R$--parity violating ones.

\nonumsection{Acknowledgments}
\noindent
D. A. D. thanks Alexander Kusenko for useful conversations.

\end{document}